\documentclass{article}

\usepackage{arxiv}

\usepackage[utf8]{inputenc} 
\usepackage[T1]{fontenc}    
\usepackage{hyperref}       
\usepackage{url}            
\usepackage{booktabs}       
\usepackage{amsfonts}       
\usepackage{nicefrac}       
\usepackage{microtype}      
\usepackage{lipsum}		
\usepackage{graphicx}
\usepackage[square,numbers]{natbib}
\usepackage{doi}

\newcommand{\trp}{\mathsf{T}}
\usepackage{siunitx}

\title{First observations of the seiche that shook the world}

\date{October 15th, 2024}	

\author{\href{https://orcid.org/0000-0003-3889-5551}{\includegraphics[scale=0.06]{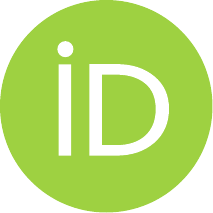}\hspace{1mm}Thomas Monahan} \\
	Department of Engineering Science\\
	University of Oxford\\
	Oxford, UK OX1 3PJ \\
	\texttt{thomas.monahan@eng.ox.ac.uk} \\
	\And
	\href{https://orcid.org/0000-0002-6365-9342}{\includegraphics[scale=0.06]{orcid.pdf}\hspace{1mm}Tianning Tang} \\
	Department of Engineering Science\\
	University of Oxford\\
	Oxford, UK \\
	\texttt{tianning.tang@eng.ox.ac.uk} \\
        \And
	\href{https://orcid.org/0000-0002-9305-9268}{\includegraphics[scale=0.06]{orcid.pdf}
        \hspace{1mm}Stephen Roberts} \\
	Department of Engineering Science\\
	University of Oxford\\
	Oxford, UK \\
	\texttt{stephen.roberts@eng.ox.ac.uk} \\
        \And
	\href{https://orcid.org/0000-0001-7556-1193}{\includegraphics[scale=0.06]{orcid.pdf}
        \hspace{1mm}Thomas A. A. Adcock} \\
	Department of Engineering Science\\
	University of Oxford\\
	Oxford, UK \\
	\texttt{thomas.adcock@eng.ox.ac.uk} \\
}

\renewcommand{\shorttitle}{\textit{arXiv} Template}

\hypersetup{
pdftitle={First observations of the seiche that shook the world},
pdfsubject={physics.geo-ph},
pdfauthor={Thomas Monahan, Tianning Tang, Stephen Roberts, Thomas A. A. Adcock},
pdfkeywords={Climate Change, Satellite Altimetry, Extreme Events, Bayesian Machine Learning},
}

\begin{document}
\maketitle

\begin{abstract}
	On September 16th, 2023, an anomalous 10.88 mHz seismic signal was observed globally, persisting for 9 days. One month later an identical signal appeared, lasting for another week. Several studies have theorized that these signals were produced by seiches which formed after two landslide-generated mega-tsunamis in an East Greenland fjord. This theory is supported by seismic inversions, and analytical and numerical modeling, but no direct observations have been made – until now. Using data from the new Surface Water Ocean Topography mission, we present the first observations of this phenomenon. By ruling out other oceanographic processes, we validate the seiche theory of previous authors and independently estimate its initial amplitude at 7.9 m using Bayesian machine learning and seismic data. This study demonstrates the value of satellite altimetry for studying extreme events, while also highlighting the need for specialized methods to address the altimetric data’s limitations, namely temporal sparsity. These data and approaches will help in understanding future unseen extremes driven by climate change.
\end{abstract}


\section{Introduction}\label{sec1}

Extreme events are evolving as a direct consequence of climate change, leading to the emergence of new, previously unobserved phenomena \cite{diffenbaugh2017quantifying, overland2022arctic}. In remote regions like the Arctic, where in-situ measurements are sparse, scientists must increasingly depend on analytical and numerical models to explore these events. However, modeling in such regions presents significant challenges due to the uncertainties in the data required to calibrate and validate these models \cite{landrum2020extremes}. Consequently, large simplifications are often necessary, resulting in substantial discrepancies between observed and modeled phenomena.
\\
\\
The mysterious 10.88 \unit{\milli\hertz} very-long-period (VLP) seismic signal, which appeared following a tsunamigenic landslide in the Dickson Fjord, Greenland, on September 16th, 2023, and the subsequent interdisciplinary scientific efforts to determine its origin, underscore these challenges. Two independent studies \cite{carrillo202416,svennevig2024rockslide} have hypothesized that the signal was driven by a standing wave, or seiche, which formed in the aftermath of the tsunami. While it is well-documented that seiches can form in resonant enclosed and semi-enclosed basins \cite{rabinovich2010seiches}, the loading-induced tilt they produce has only been observed locally ($<30$ \unit{\km}) and for short durations ($<1$ hour)\cite{svennevig2024rockslide,amundson2012observing}. Moreover, no prior evidence exists of persistent fluid sloshing (lasting several days) without an external driver. The 9-day attenuation of the globally detected VLP seismic signal is thus highly anomalous. Even more curious was the reappearance of the signal on October 11th, 2023, as noted by \cite{svennevig2024rockslide}, with approximately half the magnitude and duration of the initial event. This recurrence coincided with a second tsunamigenic landslide in the same gully in the Dickson Fjord.
\\
\\
The rock-ice avalanches and subsequent tsunamis from both events have been well-documented through a combination of satellite and field observations, including evidence of tsunami run-up within the fjord and as far away as the research station at Ella Ø, over 72 \unit{\km} away \cite{carrillo202416, svennevig2024rockslide} (Section \ref{dickson_description}, and Figure \ref{fig:fjord_overview}). In contrast, evidence for the seiche has, by necessity, relied solely on a combination of analytical and numerical models, supplemented by seismic observations.  While the seismic amplitudes are well reproduced, significant discrepancies remain between these studies in the estimated initial amplitude of the seiche (ranging from 2.6 \unit{\m} to 7.4–8.8 \unit{\m}). Empirical observations are essential not only to confirm the existence of the seiche but also to validate the models and refine our understanding of the event dynamics.
\begin{figure} 
	\centering
	\includegraphics[width=1\textwidth]{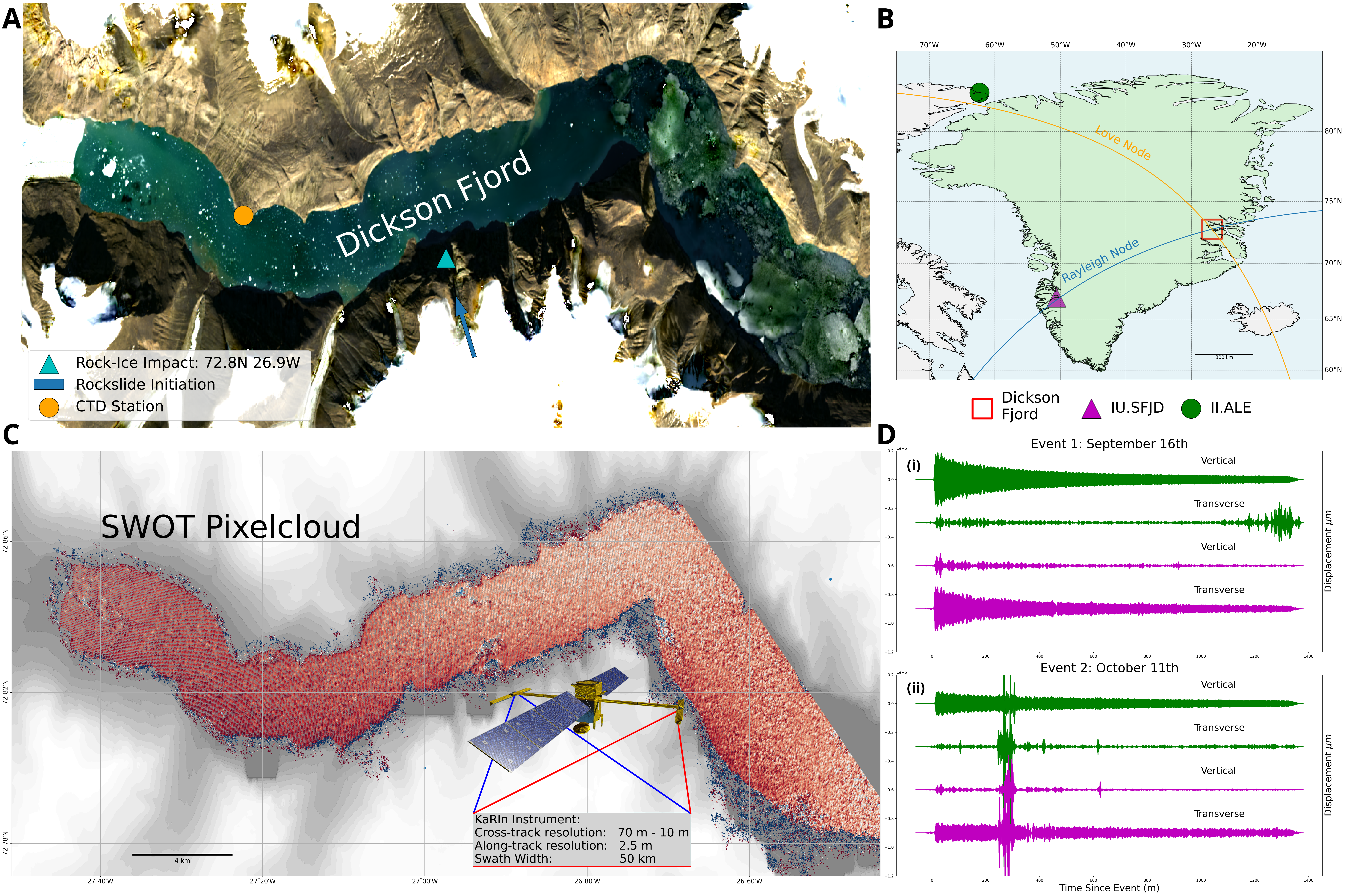} 
\caption{\textbf{Dickson Fjord study region, SWOT measurements, and in-situ measurements} (\textbf{A}) Sentinel-2 image of the Dickson Fjord in summertime, with rock-slide, CTD tide-gauge, and atmospheric station shown. (\textbf{B}) Visualization of the study region, and nearby IU.SFJD and II.ALE seismic station. Rayleigh and Love nodes are plotted in blue and orange. (\textbf{C}) SWOT pixelcloud measurements from a single pass over Dickson Fjord, measurements colored by measured sea-surface height. (\textbf{D.i \& D.ii}) Seismic observations at II.ALE (green) \cite{scripps1986global} and IU.SFJD (magenta) \cite{albuquerque1988global} bandpass filtered between 10-13mHz for the September 16th (i) and October 11th (ii) events. Time units are minutes.} 
	\label{fig:overview} 
\end{figure}
\subsection{The evidence}
To support the conclusions that the 10.88 \unit{\milli\hertz} VLP signal was produced by a seiche in the Dickson Fjord the authors of \cite{carrillo202416} and \cite{svennevig2024rockslide} applied two independent approaches. Both studies identify the observed radiation pattern of Rayleigh and Love waves to be consistent with an oscillating single force perpendicular to the Dickson Fjord. Additionally, both groups identify seismic source locations near the Dickson Fjord. Here we present a much abbreviated summary of the additional methods and evidence.
\\
\\
By performing a seismic inversion on three sets of teleseismic arrays, the authors of \cite{carrillo202416} isolate a predominantly horizontal and perpendicular force to the Dickson Fjord. This finding serves as the basis of a simple analytical model of the sloshing physics by considering a simplified rectangular fjord geometry of 2 x 20 \unit{\km}. Using the single force inversion, they identify an initial horizontal force of approximately 160 \unit{\giga\newton} which leads to an estimated initial amplitude of 2.6 \unit{\m}. While a 2-d finite difference modelling approach was applied in the simplified rectangular geometry, this model only serves to illustrate that the fundamental-mode oscillation can form. Much effort is devoted to identifying the physical drivers of the observed damping, which, due to our approach not requiring these, we do not discuss further.   
\\
\\
The large interdisciplinary team in \cite{svennevig2024rockslide} use a combination of high-resolution numerical simulation and analytical modelling to corroborate their claims. Two numerical approaches are considered, however, their preferred approach employs a nonlinear hydrostatic model implemented in HySea \cite{macias2021multilayer} which treats the landslide as a granular flow. Using an extremely fine grid spacing of 3 \unit{\m}, the simulation stabilizes into a slowly decaying seiche after approximately 5 minutes with an initial amplitude of 7.4 \unit{\m}. Notably, the first eigenmode has an oscillation frequency of 11.45 \unit{\milli\hertz} (85 seconds), which differs from the observed VLP signal frequency of 10.88 \unit{\milli\hertz} (92 seconds). This numerical simulation is then employed as a source time function to generate global seismic waveforms. Through direct comparison of the simulated envelope, the authors find good agreement between the synthetic and observed signal attenuation. An analytical model is also utilized, using a more realistic simplified geometry than in \cite{carrillo202416}. The authors identify the initial force to be 500 \unit{\giga\newton}, significantly larger than the 160 \unit{\giga\newton} estimate in \cite{carrillo202416}. This yields an estimated initial seiche amplitude of 8.8 meters. 
\\
\\
While both studies provide compelling evidence that the source of the persistent 10.88 \unit{\milli\hertz} signal was a seiche originating in the Dickson Fjord, significantly different values are obtained for the initial amplitude of the seiche (2.6 \unit{\m} vs. 7.4-8.8 \unit{\m}). Both studies attribute these discrepancies to unmodelled effects. We note that both studies consider significantly different simplified fjord geometries e.g. \cite{carrillo202416} assume a fjord width of 2 \unit{\km}, and a length of 20 \unit{\km}, and \cite{svennevig2024rockslide} assume a width of 2.7 - 2.88 \unit{\km} (depending on the figure) and a length of 10 km. Naturally, these choices will lead to different analytical estimates of seiche amplitudes. Here, we offer a completely different approach -- using the first direct empirical observation of this phenomenon to answer these questions.

\section{Results}\label{sec2}
\subsection{Surface Water Ocean Topography (SWOT) Mission}
In contrast to in-situ devices, satellite altimetry provides near-global measurements, albeit with an inevitable trade-off in temporal sampling \cite{chelton2001satellite}. After more than 30 years, these data have revolutionized our understanding of many oceanic and climatic processes \cite{abdalla2021altimetry}. However, significant challenges arise in the study of extreme events due to a combination of the temporal sparsity, and the 1-d nature of the observations. Conventional Nadir altimeters sample data directly beneath the spacecraft, producing 1-d profiles along the sea-surface. This sampling severely limits the ability to draw conclusions regarding the spatial dynamics of extremes, and often leaves events unobserved altogether. 
\\
\\
The new wide-swath Surface Water Ocean Topography mission, launched on December 15th, 2022, has overcome many of these deficiencies \cite{morrow2019global}.  Unlike conventional Nadir altimeters, the KaRIn instrument onboard SWOT provides ultra-high resolution 2-d measurements of ocean surfaces extending 50 \unit{\km} on either side of the spacecraft \cite{fjortoft2013karin}. SWOT provides high accuracy measurements directly up-to coastlines, and uniquely into Fjords, with an effective pixelcloud resolution of 6 \unit{\m} along-track and 10 \unit{\m} in the cross-track direction. An overview of these sampling characteristics and an example of a single SWOT pass over the Dickson Fjord is shown in Figure \ref{fig:overview}, Panel C. The pixelcloud contains more than 300,000 measurements and provides complete coverage of the study region.
\\
\\
After transitioning to the Science orbit phase of the SWOT mission, SWOT made several observations of the Dickson Fjord shortly after the occurrence of both tsunamigenic landslides. For the September 16th event, these passes occurred 0.5-days, 1.5-days, and 4.8-days after the VLP developed. For the October 11th event, only a single `usable' pass existed 0.5 days after the VLP began. 

\subsection{Empirical Observations}
\begin{figure} 
	\centering
	\includegraphics[width=1\columnwidth]{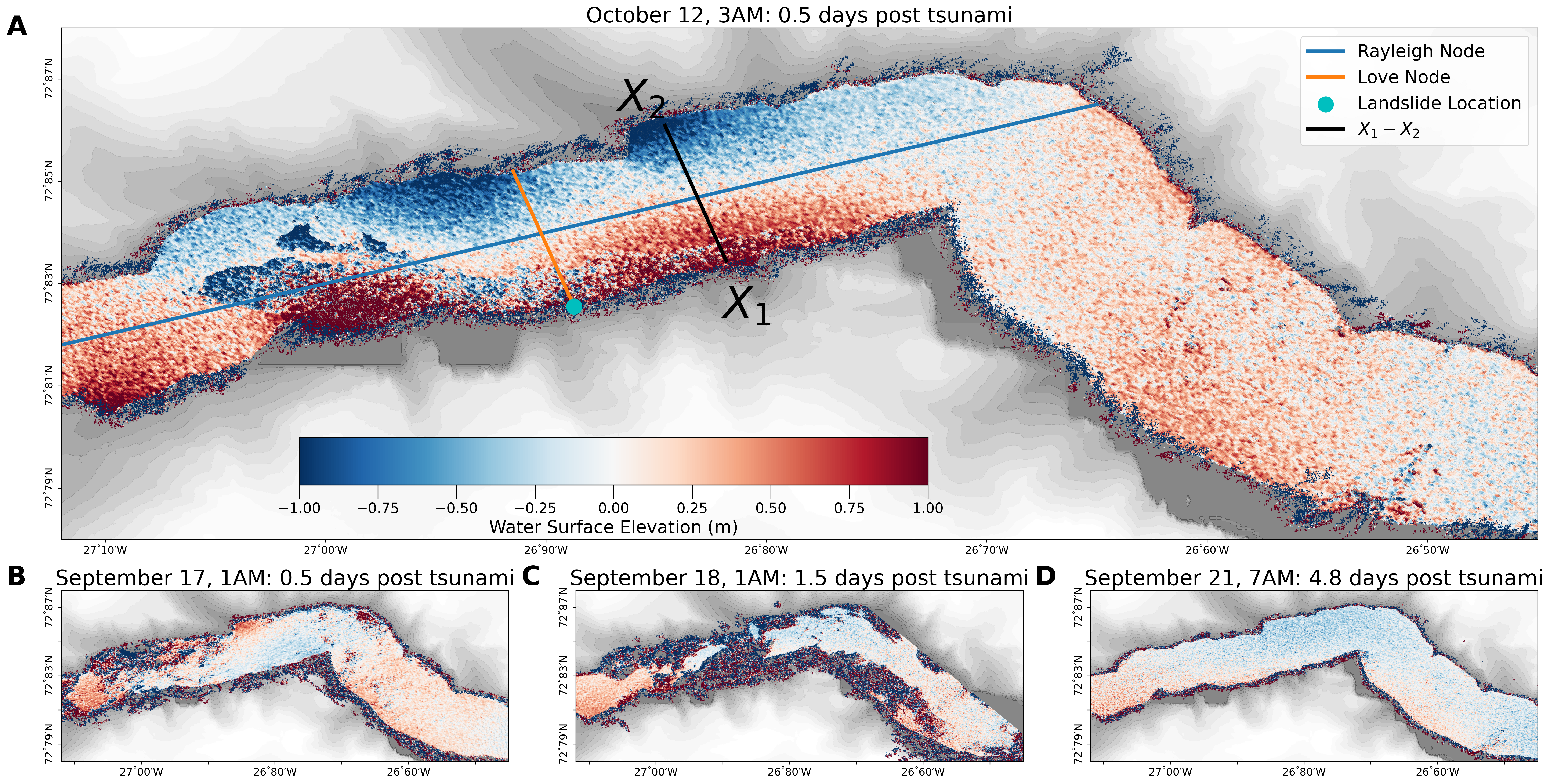} 
\caption{\textbf{Pixelcloud sea-surface elevation maps of the Dickson Fjord in the days following the two tsunamis.} (\textbf{A}) SWOT observation of the fjord 0.5 days after the October 11th tsunami. Rayleigh and Love nodes are overlaid to show the theorized axis of propagation. (\textbf{B,C,D}) Consecutive SWOT observations of the fjord 0.5 days, 1.5 days, and 4.8 days after the September 16th event respectively.} 
	\label{fig:swot_events} 
\end{figure}
SWOT pixelcloud observations of the Dickson Fjord for both the September and October events are shown in Figure \ref{fig:swot_events}. For the October 12th observation (0.5 days post-tsunami), a large negative cross-channel slope (relative to the line $\overline{X_1 X_2}$) can be observed across the minor-axis of the fjord. Here we refer to the longitudinal-axis of the fjord as the major axis with the minor axis sitting perpendicular to it. While some noise artifacts exist around 27$^{\circ}$W, the spatial distribution of resonant nodes is in good agreement with the high-resolution tsunami simulation in \cite{svennevig2024rockslide}. 
\\
\\
Unfortunately, strong noise artifacts muddle large portions of the September 17th and 18th observations. However, on September 17th (0.5 days post-tsunami), a cross-channel slope can be observed antiphase to both the October 11th and September 18th observations. Due to the noise artifacts on September 18th, estimates of the slope are not accurate. A description of the possible sources of these artifacts is given in the Section \ref{swot_data}. The September 21st observations exhibit almost no noise artifacts and a very shallow negative cross-channel slope. The spatial distribution of nodes is consistent with those observed on October 12th and the tsunami simulation in \cite{svennevig2024rockslide}. We note that significant discrepancies exist between the simulated and observed sea-surface, particularly in the upper reaches of the Fjord. Here, the SWOT observations suggest the presence of a persistent water build-up which appears to go unmodeled due to boundary choices. 
\subsection{Seismic Attribution}
\begin{figure} 
	\centering
	\includegraphics[width=1\columnwidth]{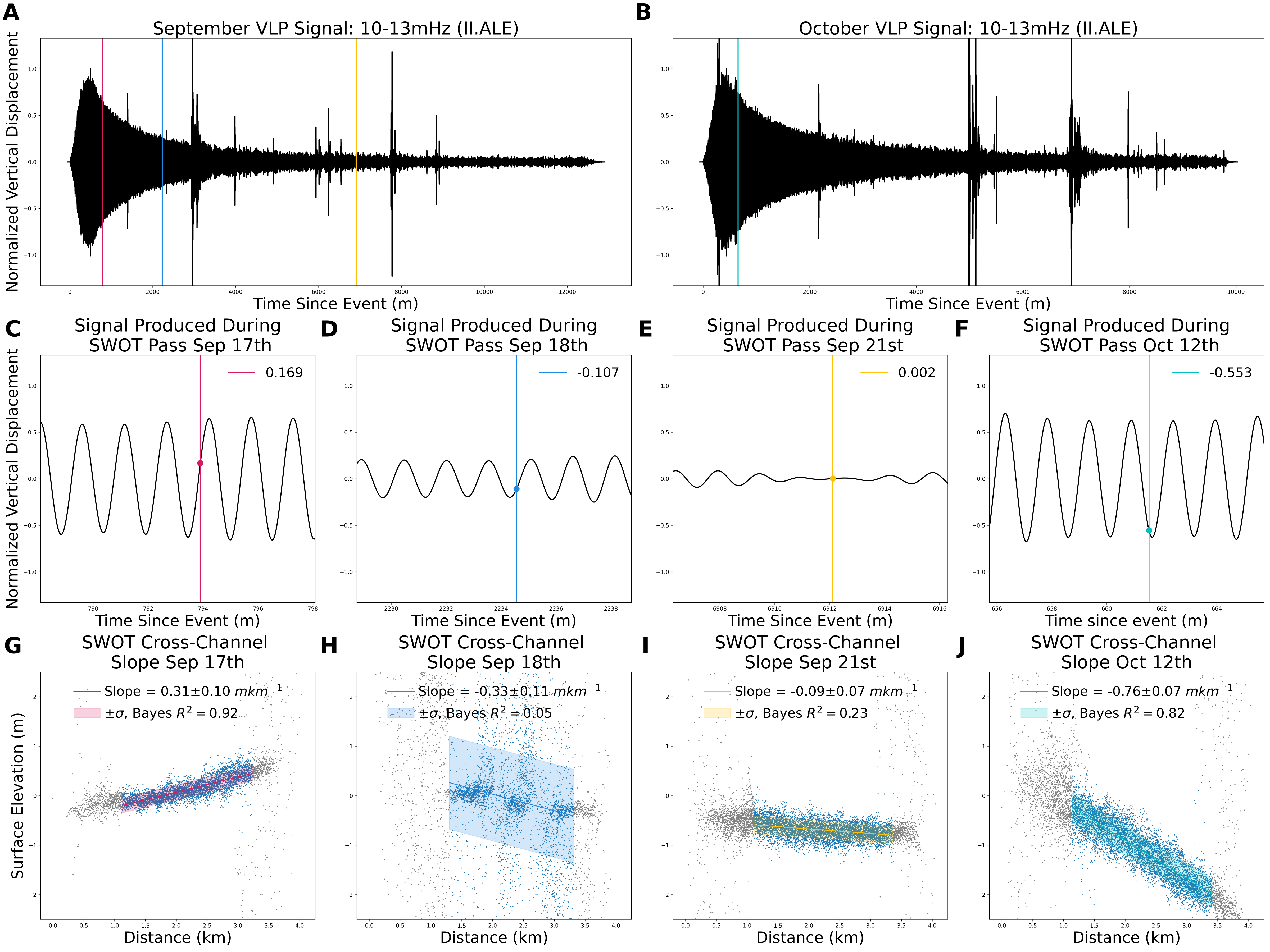} 
\caption{\textbf{Seismic observations of September and October VLP signals at II.ALE Seismic Station and SWOT cross-channel slopes.} (\textbf{A,B}) Normalized VLP signals filtered between 10-13 \unit{\milli\hertz} for the September and October events respectively. SWOT observations are given by vertical lines. (\textbf{C-F}) Normalized VLP signals with signals observed by SWOT shown as vertical lines. Observed magnitudes relative to the maximum amplitude are shown. (\textbf{H-K}) Corresponding SWOT cross-channel observations from $X_1$ to $X_2$. Slope estimates and associated Bayesian $R^2$ values from a Bayesian linear model are provided (Section \ref{Bayes}).} 
	\label{fig:seismic} 
\end{figure}
The SWOT data alone cannot estimate the total amplitude of the seiche as the observations could have occurred at any phase of the oscillation period. To overcome this, SWOT observations are referenced to seismic data from the II.ALE seismic station located at 82.5033$^{\circ}$N 62.3500$^{\circ}$W (1322.9 \unit{\km} away). The II.ALE station sits directly adjacent to the Love node and is thus characterized by almost exclusively Rayleigh waves as shown in Figure \ref{fig:overview}, Panels B, D and E. The measured vertical displacement at II.ALE is filtered between 10 and 13 \unit{\milli\hertz}, and shown for both events in Figure \ref{fig:seismic}, Panels A and B. SWOT observations are shown as vertical lines. We estimate a phase speed of 4.03 \unit{\km\per\s} using a heterogeneous Earth model \cite{pasyanos2014litho1}, and an approximate distance from the seismic source of 1409.5 \unit{\km} computed using the same heterogenous Earth model (see Section \ref{seismic_attribution} for details, and Figure \ref{fig:fjord_overview} for relative location). Estimates of the uncertainty in this value, as well as validation using synthetic data are also provided in Section \ref{seismic_attribution}. The relative magnitude and phase of the seiche can be directly determined through comparisons with the observed ground motion. Snapshots of the observed vertical VLP signal with SWOT observations highlighted are shown in Figure \ref{fig:seismic}, Panels C-F. The SWOT observed cross-channel slopes, computed between points $X_1$ and $X_2$ which gave rise to these signals are plotted directly below in Figure \ref{fig:seismic}, Panels G-J. The observed cross-channel slopes nicely correspond to the vertical displacement produced at station II.ALE. That is, negative cross-channel slopes (from $X_1$ to $X_2$) are associated with a negative vertical displacement and vice versa. The magnitudes also show good agreement. We note that this is exactly what is expected from the horizontal force produced by the seiche oscillation (Section \ref{analytical}) \cite{svennevig2024rockslide}.
\\
\\
To validate these observations, the normalized VLP signal is used to estimate the initial amplitude of the seiche. Estimated slopes are computed using a Bayesian linear model (Section \ref{Bayes}). Uncertainty estimates are obtained at both the parameter level, alongside estimates of the noise content of the data. Due to the fact that each study assumes a different width and length of the fjord, we instead consider each estimate in terms of the corresponding cross-channel slope at maximum amplitude (MXCS). For \cite{svennevig2024rockslide} their initial amplitude estimate of 7.4-8.8 \unit{\m} translates to an MXCS of 2.56-3.13 \unit{\m\per\km}, and the initial amplitude estimate of 2.6 \unit{\m} by \cite{carrillo202416} yields an MXCS of 1.3 \unit{\m\per\km}. Using the SWOT observations from September 17th we estimate the MXCS to be $1.83 \pm 0.59$ \unit{\m\per\km}. This value sits in between the estimates of both previous studies, however, the analytical estimate by \cite{carrillo202416} sits at the very bottom of the confidence intervals, and the numerical estimate from \cite{svennevig2024rockslide} just outside the top. Data from September 18th and 21st were not utilized due to large relative uncertainties and low Bayesian $R^2$ scores \cite{gelman2019r} (see  Section \ref{Bayes}). For the October event, we estimate the MXCS to be $1.37 \pm 0.13$ \unit{\m\per\km}. The fact that the October 12th observations occurred near a local minimum in the seiche's oscillation allows for tighter uncertainty estimation. 
\section{Ruling out other suspects}
While the SWOT data provides an unprecedented look at the instantaneous water levels in the Dickson Fjord, it is only that; a snapshot. The observed cross-channel slopes conform to our expectation of a standing wave oscillating perpendicular to the major-axis of the Fjord. However, there are other geophysical phenomena which can give rise to large cross-channel slopes in enclosed basins, namely tides \cite{caceres2003observations} and wind-driven circulation (Ekman transport) \cite{cottier2010arctic}. Consideration is given to each of these possible causes. 
\subsection{Tides}
\begin{figure} 
	\centering
	\includegraphics[width=1.0\textwidth]{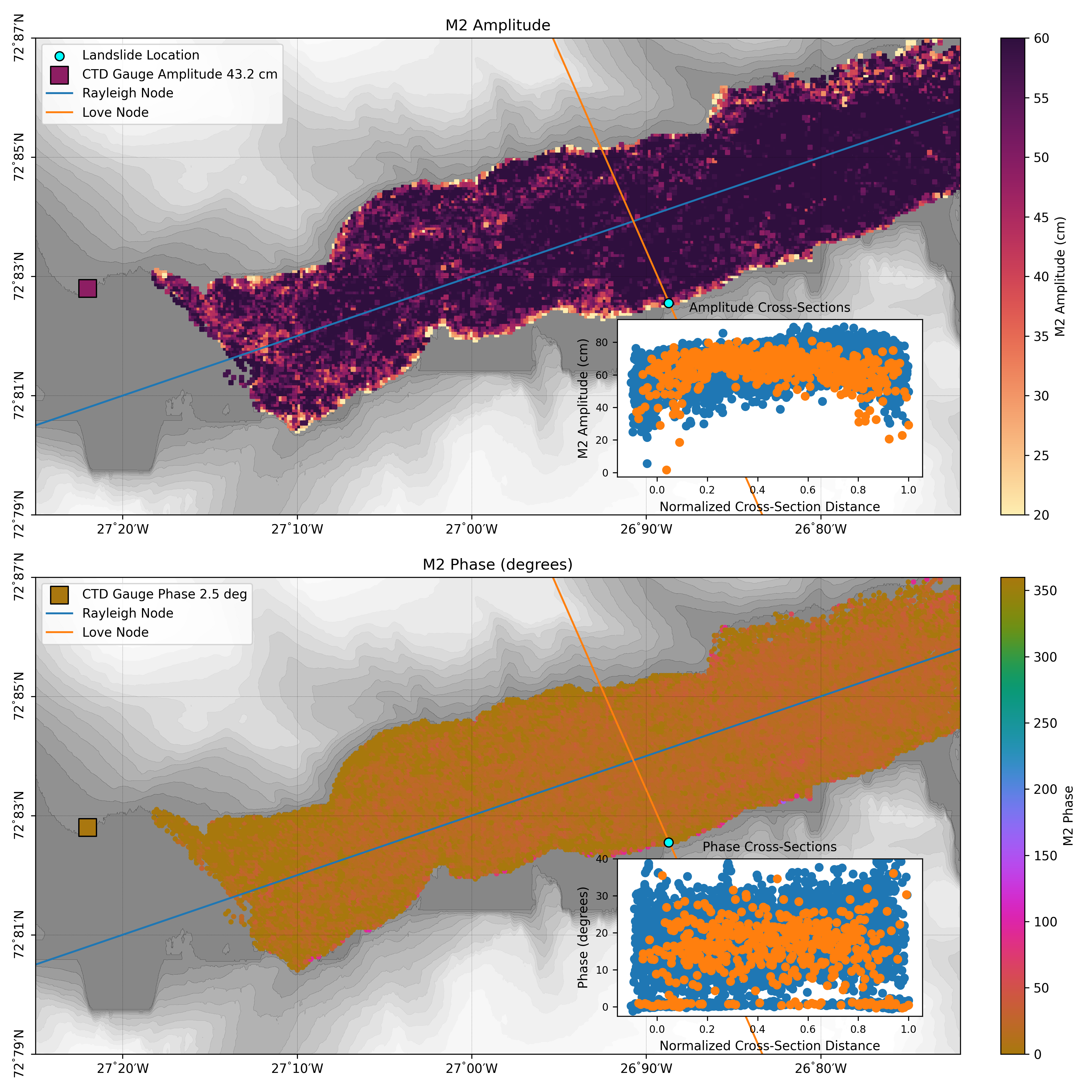} 

	\caption{\textbf{Estimates of the dominant lunar tide, M2, from SWOT pixelcloud data using a spatially coherent variational Bayesian harmonic analysis.} (\textbf{Top}) Estimated amplitudes, and (\textbf{Bottom}) corresponding phase lags. Estimates are only made for points which have at least 15 measurements. Inset plots show how the amplitude and phase vary along the rayleigh (blue, left to right) and love (orange, bottom to top) nodes respectively. The M2 amplitude and phase lag computed from the depth measurements at the CTD station are shown by the square.}
	\label{fig:tides} 
\end{figure}
While tides are obviously not the source of the approximately 92-second seismic signal, due to the sparsity of the SWOT measurements, tidally driven cross-channel slopes could lead to false conclusions about the presence of a seiche in the SWOT data. As there is only one tide gauge in the region, insufficient in-situ information exists to rule this out. Additionally, small fjords are poorly resolved by state-of-the-art global tide models due to interpolation and can thus not be relied upon \cite{hart2024tides, stammer2014accuracy}. We instead directly analyze the SWOT pixel cloud data obtained over the year following the two events using a novel spatially coherent Bayesian harmonic analysis procedure \cite{monahan2024tidal}. This procedure has been shown to improve tidal estimation over conventional least-squares approaches for extremely sparse reference series and in complex coastal regions. A complete description of our implementation is given in Section \ref{tide_estimation}. 
\\
\\
Figure \ref{fig:tides} shows empirical estimates of the amplitude and phase lag of the dominant lunar ocean tide, M2. Pixelcloud data is obtained and analyzed from October 20th, 2023 to September 20th, 2024. Due to the presence of winter sea-ice, and errors in the SWOT data, we identified only 20 usable passes. This limited data is insufficient to resolve additional tides, however, suffices for M2. As expected, both amplitude and phase exhibit small linear trends along the major-axis of the fjord and estimates are in good agreement with the CTD gauge (Figure \ref{fig:tides}). Along the cross-channel direction, a slight rise in amplitude (5-10 \unit{\cm}) is seen in the center of the Fjord but tapers off closer to the shores. This behavior may influence the curvature of the cross-channel profile, but will not induce the linear slope observed. The derived phase lags are uniformly distributed along the minor-axis of the fjord. Due to the fact that cross-channel variations in phase lag give rise to cross-channel slopes, we conclude that the observed slope in the SWOT data is not tidally driven. 
\subsection{Ekman Transport}
\begin{figure} 
	\centering
	\includegraphics[width=1.0\textwidth]{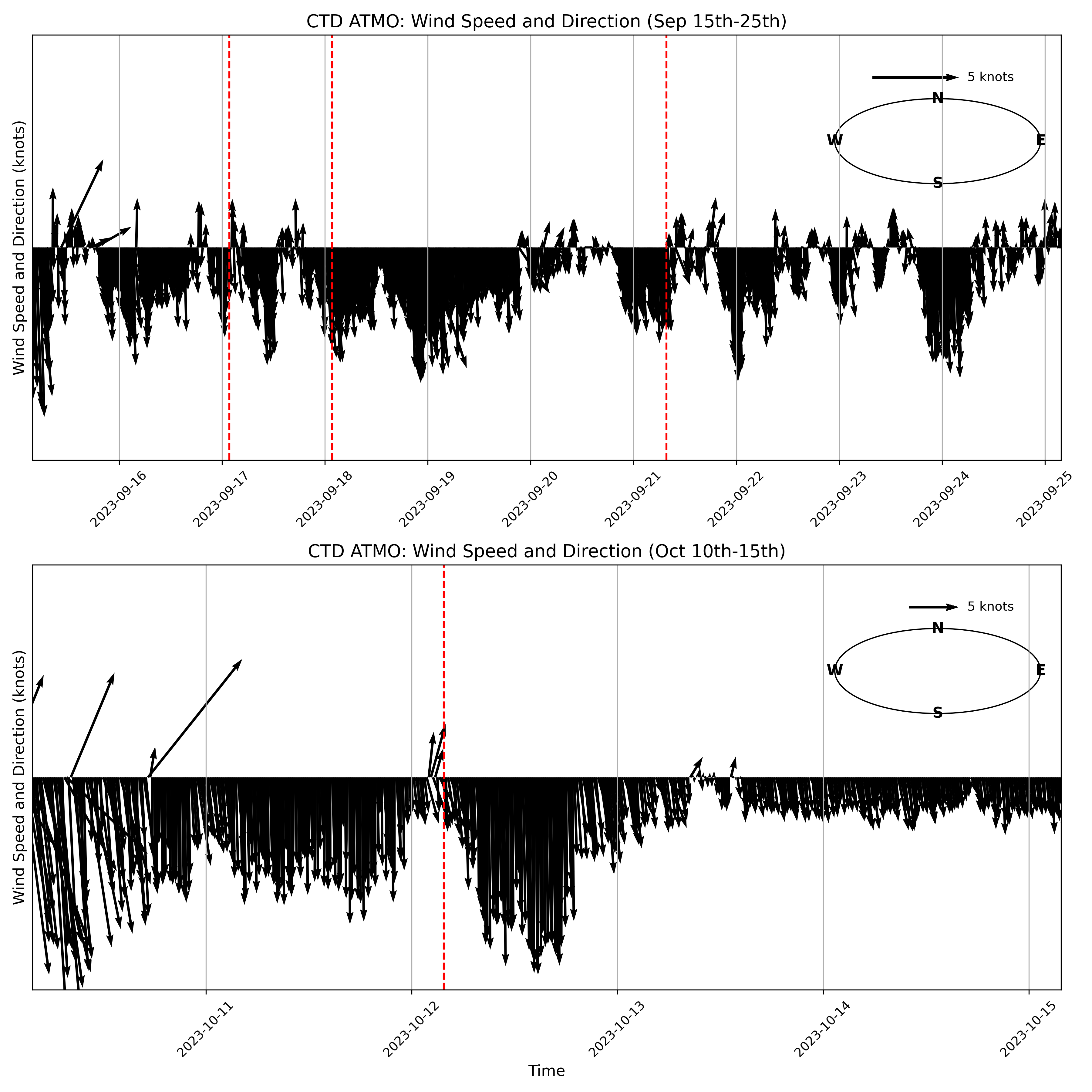} 
	\caption{\textbf{Wind speed and direction from the Dickson Fjord CTD station during each event}(\textbf{Top}) Wind speed and direction over the duration of the September 16th VLP signal. (\textbf{Bottom})  Wind speed and direction over the duration of the October 11th VLP signal.}
	\label{fig:ekman1} 
\end{figure}

Ekman transport occurs as a consequence of sustained wind-stress \cite{price1987wind}. In the Northern hemisphere, the coriolis effect causes water to propagate $90^{\circ}$ clockwise to the incident wind direction. We evaluate the wind-speed and direction from the CTD atmospheric station shown in Figure \ref{fig:overview} Panel A. Results are shown in Figure \ref{fig:ekman1}. Observations over the duration of the first VLP signal in September show that all SWOT measurements occur after sustained periods of southerly winds at approximately 5 knots. This wind-stress can induce a build-up of water in the westerly direction, but should not induce a cross-channel slope. The observations on October 12th follow yet another sustained period of Southerly wind, this time with a higher magnitude of approximately 10 knots. The SWOT observations occur as the wind is switching direction at low magnitudes ($< 5$ knots). While sustained winds in the North-Western direction could give rise to cross-channel slopes, the low magnitude and short-duration of this change are unlikely to give rise to the large (2 m) cross-channel slope observed. 

\section{Discussion}\label{sec12}

Our study provides the first direct observational evidence of the seiche in the Dickson Fjord. Based on the seismic attribution, and systematic ruling out of other dynamic phenomena, we conclude that the observed variability in the SWOT data is consistent with that of a slowly decaying seiche. Thus, this study corroborates the numerical and analytical evidence given in \cite{carrillo202416} and \cite{svennevig2024rockslide}, that the globally observed VLP signal which originated on September 16th, 2023, was due to a seiche which formed after a megatsunami. Additionally, we conclude that the smaller VLP signal observed on October 11th, was also a seiche originating from a smaller tsunami in the very same fjord. Notably, this constitutes the first ever empirical observation of a short period seiche persisting for several days without an external driver. 
\\
\\
Using an empirical and completely independent approach, we estimate the maximum cross-channel slope (MXCS) of the September seiche to be $1.83 \pm 0.59$ \unit{\m\per\km}. This value falls between the analytical estimate of \cite{carrillo202416} at 1.3 \unit{\m\per\km} and the numerical/analytical estimates of \cite{svennevig2024rockslide}, which range from 2.56 to 3.13 \unit{\m\per\km}. Due to the relatively large uncertainty, our empirical estimate--when considered in isolation--provides limited insight into the true initial value. However, our analysis of the October VLP signal produced a first estimate of the October seiche's MXCS at $1.37 \pm 0.13$ \unit{\m\per\km}. The smaller uncertainty in this case, owing to the October 12th observation occurring near a local minimum of the vertical displacement, gives greater confidence in this estimate.
\\
\\
Seismic data from the II.ALE and IU.SFJD stations show that the initial magnitude of the October VLP signal was approximately twice that of the September event (3 \unit{\micro\m} vs. 1.5 \unit{\micro\m} at II.ALE). Since the horizontal force is proportional to the average cross-channel slope (see Section \ref{analytical}), we made a second estimate of the September MXCS to be $2.74 \pm 0.26$ \unit{\m\per\km}. This value closely aligns with the numerical/analytical estimates of 2.56–3.13 \unit{\m\per\km} reported by \cite{svennevig2024rockslide}. Therefore, based on the relative magnitudes of the two forces at adjacent seismic stations, and the robust estimate of the October 12th event, we conclude that the numerical and analytical estimates by \cite{svennevig2024rockslide} are in good agreement with the real data. Furthermore, we argue that the estimate provided in \cite{carrillo202416} likely underestimates the true magnitude due to inaccuracies in the assumed Fjord geometry, and the underestimation of the initial force at only 160 GN. If we consider a mean fjord width of 2.88 \unit{\km} as in \cite{svennevig2024rockslide}, our empirical estimate suggests the tsunami stabilizes into an initially 7.9 meter tall seiche.

\section{Conclusion}\label{sec13}

This study highlights the value of wide-swath satellite altimetry in characterizing extreme events. As shown, these data provide the opportunity to connect and understand the complex interactions between climate change and the different components of the geosphere. However, this work also emphasizes the importance of specialized, interdisciplinary methods to address the intrinsic limitations of these data, particularly the challenges posed by temporal sparsity. As noted, the SWOT data at its present level of processing, is not trivial to work with. Dedicated efforts are needed to improve reprocessing of these data in fjords. Additionally, open-source tools that bridge the gap between raw data and analysis pipelines are essential for enabling non-expert users to utilize these data.
\\
\\
\noindent While sufficient observational and bathymetric data existed to recreate the observed seiche dynamics numerically, for many remote regions this is not the case \cite{landrum2020extremes}. Indeed, while the effects of climate change are felt globally, the largest and fastest changes are often in these regions \cite{rantanen2022arctic}. As such, while we echo the claims from previous authors about the need for more in-situ sea-level gauge sensors, we believe a follow-on to the SWOT mission and investment in future wide-swath and non-sun-synchronous altimetry missions is also critical to monitoring these effects. Furthermore, we stress that a key tool for identifying these events is the accurate computation of sea-level anomalies by applying different geophysical corrections. Due to the complex and narrow geometries of fjords, and a lack of historical altimetric measurements, tidal estimates in these regions are poor \cite{hart2024tides}. This is yet another area where SWOT and future wide-swath altimetry missions can reduce these deficiencies. 

\section{Methods}\label{sec11}
\subsubsection{SWOT data and Processing}
\label{swot_data}
\begin{figure} 
	\centering
	\includegraphics[width=1.0\textwidth]{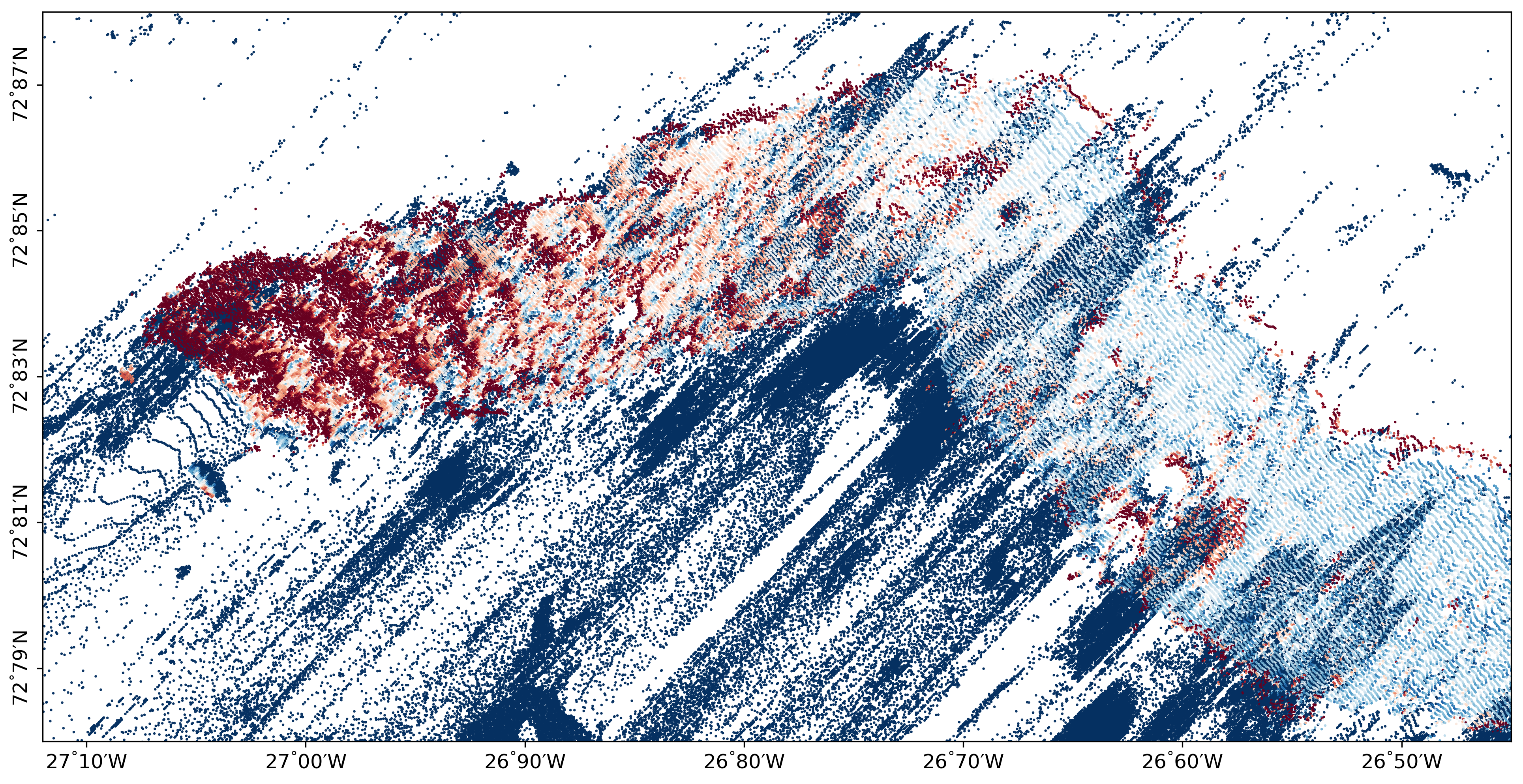} 
	\caption{\textbf{Example SWOT observation with noise contamination.} Observation produced by data file: $SWOT\_L2\_HR_PIXC\_005\_292\_021R\_20231023T020832\_20231023T020843\_PGC0\_01.$ }
	\label{fig:bad_obs} 
\end{figure}

\begin{figure} 
	\centering
	\includegraphics[width=1.0\textwidth]{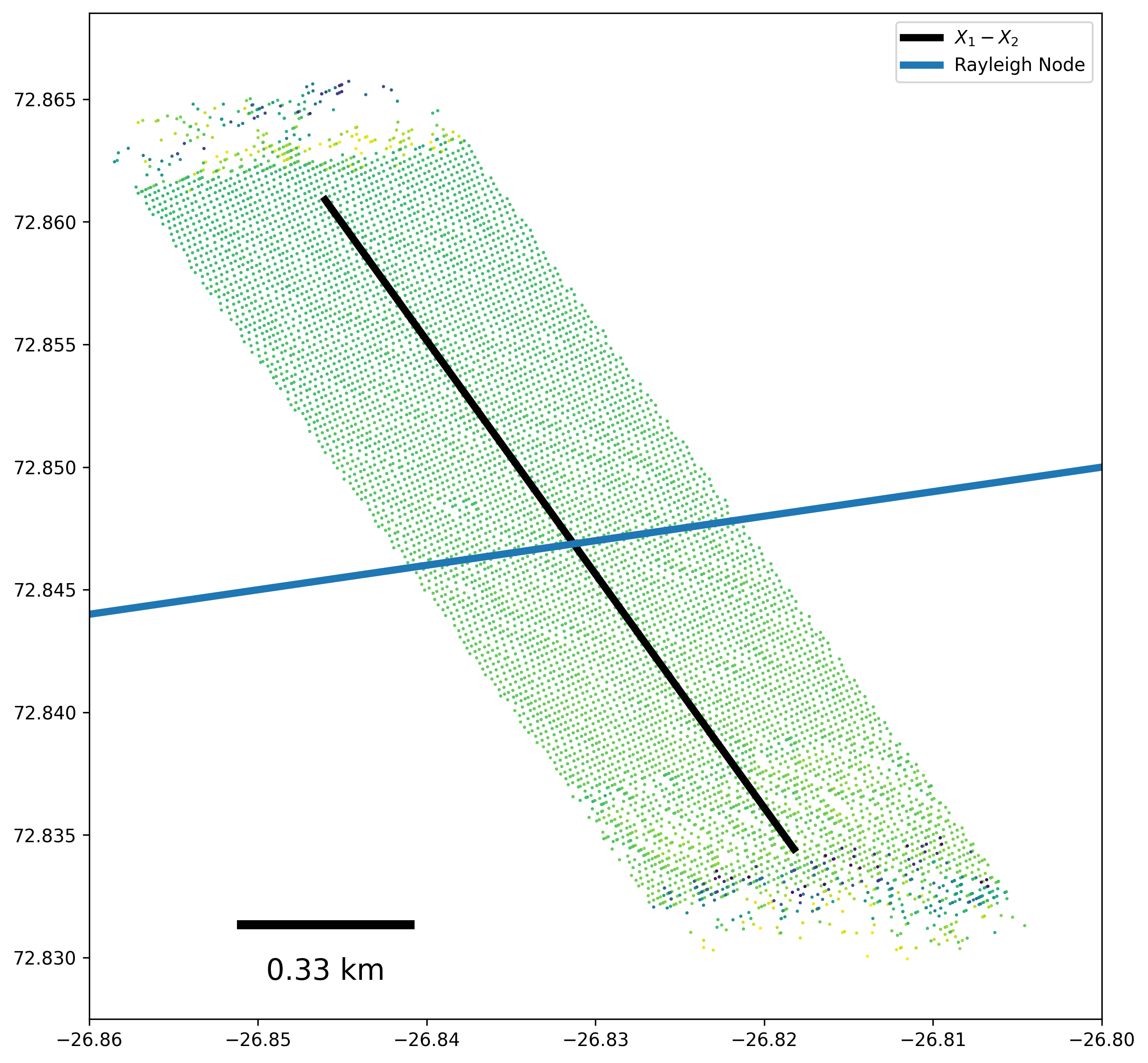} 
	\caption{\textbf{Example SWOT cross-section used to compute the cross-channel slope}. SWOT observations are colored according to sea-surface height.}
	\label{fig:cross_section} 
\end{figure}

Pixelcloud data from the SWOT mission are obtained through open Earth Access (last accessed on October 3rd, 2023). We utilize the version 2.0, HR pixecloud data with shortname ``SWOT\_L2\_HR\_PIXC\_2.0'' in the earthaccess API. At the time of access, this constitutes the highest processing level available. The Dickson Fjord landslides in September and October of 2023 occurred shortly after the transition of SWOT to the Science phase. The SWOT Science phase is characterized by a 20.86 day repeat orbit with 10-day sub-cycles. The orbit is at an inclination of 77.6$^\circ$, and is thus non-sun-synchronous, which reduces tidal aliasing. A consequence of the 10 day sub-cycles is that repeat measurements occur in groups leading to relatively short measurement gaps at some portions of the cycle, and complete gaps in the other. Data for the September and October events were obtained by filtering all available passes between September 16-26, and October 10-18th respectively. We select all passes which fall into the bounding box given in Table \ref{tab:bounding_box}. For the September period, this yields 5 initial passes occuring 0.5 days, 2 days, 3 days, 4 days and 5 days after the event respectively. Similarly, for the October period, 2 initial passes are identified at 0.5 days and 6 days post-tsunami. We process the SWOT data in two stages, first through manual inspection and then using a standard preprocessing procedure.
\\
\\
\textbf{Manual Inspection:} Prior to processing each pass we perform a manual inspection. Due to limitations in the L2 processing algorithms applied to SWOT data, entire passes can be contaminated to the point where the data becomes unusable. An example of this contamination can be seen in Figure \ref{fig:bad_obs}. These errors are visually obvious and are manually flagged and removed through visual inspection. This yields a final set of observations for the September event 0.5 days, 2 days, and 5 days, and only a single observation for the October event taken 0.5 days post-tsunami. 
\\
\\
\textbf{Data Processing:} All standard geophysical corrections are applied including wet and dry troposphere delays and cross-over corrections. Unlike the Cal/Val SWOT orbit, the cross-over correction for the Science phase is much more accurate due to the significant reduction in time between crossovers. Geoid corrections are applied based on the EGM2008 geoid model. We do not apply the included FES2022 tidal corrections as they are interpolated in Fjord regions. A comprehensive discussion on how tides are dealt with is given in Section \ref{tide_estimation}. After applying the geophysical corrections, we filter out all measurement values with sea-surface heights exceeding $\pm 4$ \unit{\m}. To obtain perceptually uniform gradients which can be compared between passes, we set the center of each colormap to be the midpoint of the cross-channel slope. Due to inaccuracies with the default tidal corrections, we found this procedure to be more robust.
\\
\\
\textbf{Cross-Channel Slope Computation:} Due to the presence of noise artifacts in some of the SWOT measurements, we rely on estimates of the cross-channel slope between the points $X_1$ and $X_2$ for our analysis. We note that the slice between these points is perpendicular to both the long-axis of the Fjord, and the Love node shown in Figure \ref{fig:swot_events}. As a consequence of the nonuniform sampling of the pixelcloud data, it is necessary to define a ``tolerance'' which defines how far a point can be from the defined cross-section to be included. We experiment with several tolerance values but converge to a value of 333 \unit{\m} on either side of the defined cross-section. An example of this is shown in Figure \ref{fig:cross_section}, with approximately 8000 measurements shown. This choice provides a representative 666 \unit{\m} swath ($\approx 5$\% of the Fjord length) with which to estimate the variability/uncertainty of the cross-channel slope. A description of the Bayesian linear model employed for this procedure is given in Section \ref{Bayes}. 
\\
\\
\textbf{SWOT Data For Tidal Estimation:} Pixelcloud data is also used to estimate the M2 tide throughout the Dickson Fjord. Due to the severe aliasing produced by the irregular temporal sampling, we make use of all available data. The presence of winter sea ice creates further difficulties as the inclusion of these data will severely degrade the accuracy of tidal estimation. As before, we locate all passes which intersect with the study region (\ref{tab:bounding_box}) between October 20th 2023, and September 10th, 2024. This yields an initial dataset of 172 passes. After manual filtering and removal of passes containing sea-ice, this leaves only 20 high-quality passes which are employed for tidal estimation. We apply the same geophysical corrections as before. In order to obtain time-series which can be fed to our spatially coherent Bayesian harmonic analysis we bin the nonuniformly sampled pixelcloud data into a fixed grid with a 50 \unit{\m} x 50 \unit{\m} resolution. Data points in each bin are averaged. The total number of data per bin is variable, thus, we restrict our analysis to bins which contain at least 15 measurements. This threshold is empirically defined based on the fact that the usage of fewer than 15 measurements in testing leading to poor tidal estimation in some regions. 
\subsubsection{Seismic Data}
Seismic data are accessed using the Federation of Digital Seismic Networks (FDSN) web service client for Obspy \cite{krischer2015obspy}. As both \cite{carrillo202416} and \cite{svennevig2024rockslide} provide a comprehensive analysis of the long-period seismic signal, we choose to focus on two representative stations: II.ALE \cite{scripps1986global}, and IU.SFJD \cite{albuquerque1988global}. In order to isolate the energy associated with the VLP, data are shown after being bandpassed between 10 and 12 \unit{\milli\hertz}. All additional filtering parameters needed to replicate our results are provided in the ``seismic\_attribution.ipynb'' notebook.   

\subsubsection{Seismic Attribution}
\label{seismic_attribution}
Seismic attribution is performed using data from the II.ALE seismic station. The station is situated 1322.9 \unit{\km} away from the landslide source at 82.5033$^{\circ}$N 62.3500$^{\circ}$W. The station sits directly adjacent to the Love node (160$^\circ$) which runs perpendicular to the major-axis of the fjord (Figure \ref{fig:overview} Panel B). As a consequence, this station receives almost exclusively Rayleigh waves which is reflected by the dominant vertical component of the VLP signal shown in Figure \ref{fig:overview} Panel D. Referencing the SWOT observations to these waveforms requires the accurate identification of the phase velocity and source location of the VLP signal. 
\\
\\
\textbf{Phase Velocity:} To compute the Rayleigh wave phase velocity we utilize a heterogeneous Earth model as in \cite{svennevig2024rockslide}. To obtain the phase velocity of the 10.88 \unit{\milli\hertz} signal we interpolate from the 10 \unit{\milli\hertz} and 15 \unit{\milli\hertz} LITHO1.0 velocity models \cite{pasyanos2014litho1}. Integrating over the path between the Dickson Fjord and the II.ALE station we obtain a mean phase velocity of 4.0393 \unit{\km\per\s}. 
\\
\\
\noindent \textbf{Source Location:} Due to uncertainties in the computed phase velocity, the source location of the VLP signal is not necessarily given by the center of the Dickson Fjord. Indeed, both \cite{carrillo202416} and \cite{svennevig2024rockslide} identify source locations which are nearby, but away from the fjord itself. Using the same heterogeneous Earth model and a Fast Marching method for beamforming, a source location 92.4 \unit{\km} from the landslide location at 72.2$^{\circ}$N 25.1$^{\circ}$W is identified \cite{svennevig2024rockslide}. This source location is approximately 1409.5 \unit{\km} away from the II.ALE seismic station. Combined with the computed phase velocity, this yields a travel time of $348.97$ \unit{\s}. 
\\
\\
\noindent \textbf{Validation With Synthetic Observations:} To test our hypothesis that the phase velocity and source location obtained using the heterogeneous Earth model are appropriate for seismic attribution we compare our approach to an independent forward modelling exercise carried out in \cite{svennevig2024rockslide}. Using their 3 \unit{\m} HySEA simulation of the seiche as a source time function, synthetic 3-component Green's functions are computed using the Syngine web service and convolved to approximate the displacement signals at the II.ALE station. In order to align the two signals, the authors identify an approximately 350 \unit{\s} shift empirically. This value is in agreement with the $348.97$ \unit{\s} travel-time computed using the heterogeneous Earth model and beam forming. 
This agreement between two completely independent methods confirms the validity of the $348.97$ \unit{\s} rayleigh wave travel time to II.ALE for seismic attribution. 
\\
\\
\noindent \textbf{Seismic Uncertainties:} While our estimated Rayleigh wave travel time is in good agreement with the empirical estimate by \cite{svennevig2024rockslide}, this represents an important source of uncertainty in the estimation of the seiche amplitude. No comprehensive approach exists for the quantification of Rayleigh wave phase velocity or beamforming location uncertainties \cite{xu2022estimation}. As such, we provide a lower-bound on our uncertainty estimate using the discrepancy between our estimate of 348.97 \unit{\s} and the empirical estimate of 350 \unit{\s} by \cite{svennevig2024rockslide}. If we assume a 92-second period of oscillation, this yields an approximate error of 2.12 \%.     

\subsubsection{Bayesian Regression}
\label{Bayes}
Both cross-channel slope and tidal estimation are performed using a Bayesian linear model. Our selection of a Bayesian approach reflects our objective to accurately quantify the uncertainty in the estimated parameters and the SWOT data themselves. Unlike conventional least-squares estimation and other frequentest variants which provide point estimates of the parameters, a Bayesian approach considers (and computes with) the probability distributions of the parameters. By representing our parameters of interest as probability distributions, the uncertainty associated with parameters in our model is explicitly represented. As will be shown, the selection of appropriate priors can yield further advantages such as natural parameter shrinkage and increased robustness to noise. 
\\
\\
Consider the linear model given by $y_i = w^\trp x_i + \epsilon_i$. Here, $y_i$ is the $i^{th}$ observation, $w$ are the estimated weights, $x_i$ is the $i^{th}$ row of an $M \times N$ design matrix of basis functions, and $\epsilon_i$ is the residual. An overview of the design matrix employed for tidal estimation is provided in Section \ref{tide_estimation}.
\\
\\
\noindent \textbf{Cross Channel Slope Estimation:} To estimate the cross-channel slope we perform a standard Bayesian linear regression. The the $i^{th}$ row of the design matrix $X$ is simply given by $X_i = [1, x_i]$, where the entry $1$ corresponds to the bias (intercept), and $x_i$ the distance along the line $\overline{X_1 X_2}$. 
\\
\\
\noindent \textbf{Bayesian Analysis:} Here we provide a brief overview of the Bayesian model utilized and the variational inference method. A complete derivation of variational inference can be found in \cite{fox2012tutorial} and a more detailed exposition of the Bayesian linear model, used here, can be found in the appendix of \cite{roberts2013astrophysically}. The basis of our analysis is Bayes' theorem, which provides a framework for updating our prior beliefs $p(\theta)$ about the distribution of all parameters (and hyperparameters\footnote{A \emph{hyper-parameter} is itself a parameter, that governs the probability distribution of another \emph{parameter}. For example, the mean and variance of a Gaussian are the hyper-parameters which define the distribution over another variable. By extension, a hyper-hyper-parameter describes a variable that controls the distribution over a hyper-parameter.}), based on a set of observations $Y = [y_0,y_1,\dots,y_N]^T$ and the design matrix $X$. This yields a posterior distribution which describes our final beliefs over $\theta$ and is given by
\begin{equation}
\label{eq:bayes}
    p(\theta | Y) =  \frac{p(\theta)p(Y|X,\theta)}{p(Y)},
\end{equation}
where $\theta$ is the set of parameters and hyper-parameters of the model, $p(\theta)$ is the prior, $p(Y|X,\theta)$ the (data) likelihood, and $p(Y)$ the marginal likelihood which acts as a normalising term in the inference (as it does not depend upon $\theta$). Here we provide a brief overview of the different components of Bayesian analysis. 
\\
\\
\noindent \textbf{Likelihood:} The likelihood $p(Y|X,\theta)$ describes the \emph{likelihood} that the observed data occurred, given our model. Here, the model is defined by both the design matrix, and our prior assumptions about the parameters $\theta$. We take the likelihood term to be of Gaussian form, equivalent to a least-squares error assumption. In our analysis, we weight the squared residual between observed $y_i$ and model prediction, by a hyper-parameter $\beta$, which represents the \emph{precision}, or inverse (co)variance of the noise. This has the effect of weighting how tightly our model should fit the data based on how ``noisy'' the data is. This assumption yields a Gaussian likelihood term of the form:
\begin{equation}
        p(Y|X,\theta) = p(Y|X,w,\beta)
        =  \left ( \frac{\beta}{2 \pi} \right )^{N/2} \exp \left \{ -\frac{\beta}{2} E_Y(w) \right \}
\end{equation}
where we see the likelihood only depends upon $w$ and $\beta$. The error functional $E_y$ is given by 
\begin{equation}
        E_Y = \sum_{i=0}^N (y_i - w^\trp x_i)^2. 
\end{equation}
\noindent \textbf{Priors:} Central to Bayesian inference is the usage of priors. Priors are distributions over the parameters included in a model, and reflect our initial expectation of the functional forms and values the parameters should have. Here we describe the choice of these priors for both parameters and hyperparameters, and describe how they impact the resultant model. The prior over all parameters $\theta$ can be factorized as 
\begin{equation}
    p(\theta) = p(w|\alpha)p(\alpha)p(\beta),
\end{equation}
where $\alpha = {\alpha_k}$ is a set of hyper-parameters which governs the scale of the multi-variate Gaussian over the weights $w$. We now treat each term individually.
\\
\\
The model weights $p(w|\alpha)$ come from a zero-mean Gaussian prior, with precision (inverse variance) $\alpha$. This choice serves two purposes. First, a Gaussian allows for weights to be either positive or negative, and is thus unbiased in this way. Second, weights will only be significantly non-zero if the data requires it. Conventionally this is referred to as an Automatic Relevance Determination (ARD) prior as it induces shrinkage over the model weights $w_k$ which do not significantly aid the model in fitting the data. Using this, for an individual weight $w_k$, the prior has the form
\begin{equation}
    p(w_k | \alpha_k) = \left(\frac{\alpha_k}{2\pi}\right)^{1/2}\exp \left \{ -\frac{\alpha_k}{2} w_k^2 \right \}.
\end{equation}
\\
\\
The set of weight precisions $\alpha$, which govern the scale of the weights $w$, are drawn from a Gamma distribution, which models the distribution over non-negative precisions.  Uninformative mixing hyper-hyper-parameters $a_0 = 10^{-2}$ and $b_0 = 10^{-4}$ are selected to yield a vague prior over each $\alpha_k$ defined as 
\begin{equation} \label{eq:alpha}
    p(\alpha_k) = \Gamma (\alpha_k ; a_0, b,0).
\end{equation}
A \emph{vague} or \emph{uninformative} prior simply means the assumed distribution is broad. This imposes minimal assumptions regarding the parameter values, whilst still providing natural parameter shrinkage \cite{ruanaidh2012numerical}. 
\\
\\
The noise precision $\beta$ (inverse variance) of the residual $\epsilon$ is also modeled as a hyperparameter within the model. Given the least-squares assumption of a Gaussian residual, we once again adopt a Gamma prior over $\beta$. These values of the mixing parameters are identical to those used in Equation \ref{eq:alpha}, but are defined by hyper-hyper-parameters $c_0 = 10^{-2}$ and $d_0 = 10^{-4}$ such that 
\begin{equation} \label{eq:beta}
    p(\beta_k) = \Gamma (\beta_k ; c_0, d,0).
\end{equation}

\noindent \textbf{Initialization:} Models are initialized using a maximum-likelihood (ML) solution such that
\begin{equation}
    w_{ML} = X^\trp Y(X^\trp X)^{-1}.
\end{equation}
The ML solution is then used to initialize the residual precision hyper-parameter, $\beta$, such that:
\begin{equation}
\beta^{-1} = \frac{1}{N} \sum_{i=1}^N (y_i - w_{ML}^\trp x_i)^2
\end{equation}

\subsubsection{Variational Inference}
Fully Bayesian solutions are obtained by marginalizing over the posterior distributions of the parameters. The difficulty arises when computing the posterior distribution, which analytically is almost always intractable. Hence, sample based approaches such as Markov-Chain Monte Carlo are often employed. While these methods are extremely good at approximating the true posterior, they scale poorly with the number of parameters included. Further, convergence is not easily assessed. Here, we adopt an approximate inference approach, called \emph{variational Bayes}, referred to herein as VB. The objective of our analysis is to infer the distributions over the individual elements of $\theta$. The basic idea of VB is to adopt analytical approximations for each distribution which can be optimized in an iterative and computationally tractable way. We first introduce an \emph{approximate posterior} $q(\theta | Y)$. The functional form of this posterior is chosen to be conjugate with the prior over $\theta$ such that $q(\theta | Y)$ factorizes as 
\begin{equation}
\label{eq:mf_approx}
    q(\theta|Y) = q(w|Y)q(\alpha|Y)q(\beta|Y).
\end{equation}
Our objective in VB is to minimize the difference between the approximate posterior $q(\theta|Y)$, and the true posterior $p(\theta|Y)$. This difference can be assessed by considering our observable, the data evidence $p(Y)$. Using our approximate posterior, we can rewrite the log evidence $p(Y)$ as the sum of two separable terms such that 
\begin{equation}
    \label{eq:fundamental}
    \log p(Y) = F \left ( p(\theta | Y), q(\theta | Y) \right ) + \mbox{KL}\left(p(\theta | Y),q(\theta | Y)\right).
\end{equation} 
This is the fundamental equation of VB and is composed of two terms. The first term is the negative variational free energy, referred to as the evidence lower bound (ELBO). This provides a strict lower bound on the model evidence. The second term is the Kullback-Liebler (KL) divergence between the approximate and true posteriors over $\theta$. This term provides natural model shrinkage as it increases with the number of free parameters $\theta$. It can be seen that maximizing $F(p,q)$ will result in the approximate posterior being as close as possible to the true posterior. Due to the fact that $q(\theta|Y)$ can be factored as Equation \ref{eq:mf_approx}, $F(p,q)$ can be maximized by iteratively optimizing each of $q(\theta|Y), q(\alpha|Y), q(\beta |Y)$ separately. Update equations for this procedure can be found in \cite{penny2002bayesian}. An implementation of this approach can be found in the replication notebooks ``Fjord\_Tides.ipynb'' and "seismic\_attribution.ipynb". 
\\
\\
\noindent \textbf{Bayesian R-squared} To evaluate the quality of the Bayesian regression we utilize the Bayesian $R^2$ proposed in \cite{gelman2019r}. This is necessary as the variance of the predicted values, can be greater than the variance of the data, thus rendering the conventional $R^2$ definition nonsensical. The modified Bayes $R^2$ is simply given by
\begin{equation}
    \text{Bayes } R^2 = \frac{\text{Var(predicted)}}{\text{Var(predicted) + Var(residual)}},
\end{equation}
where $\text{Var(residual)}$ is the expected variance of the errors as given by the model.

\subsubsection{Tidal Estimation}
\label{tide_estimation}
\begin{figure} 
	\centering
	\includegraphics[width=1.0\textwidth]{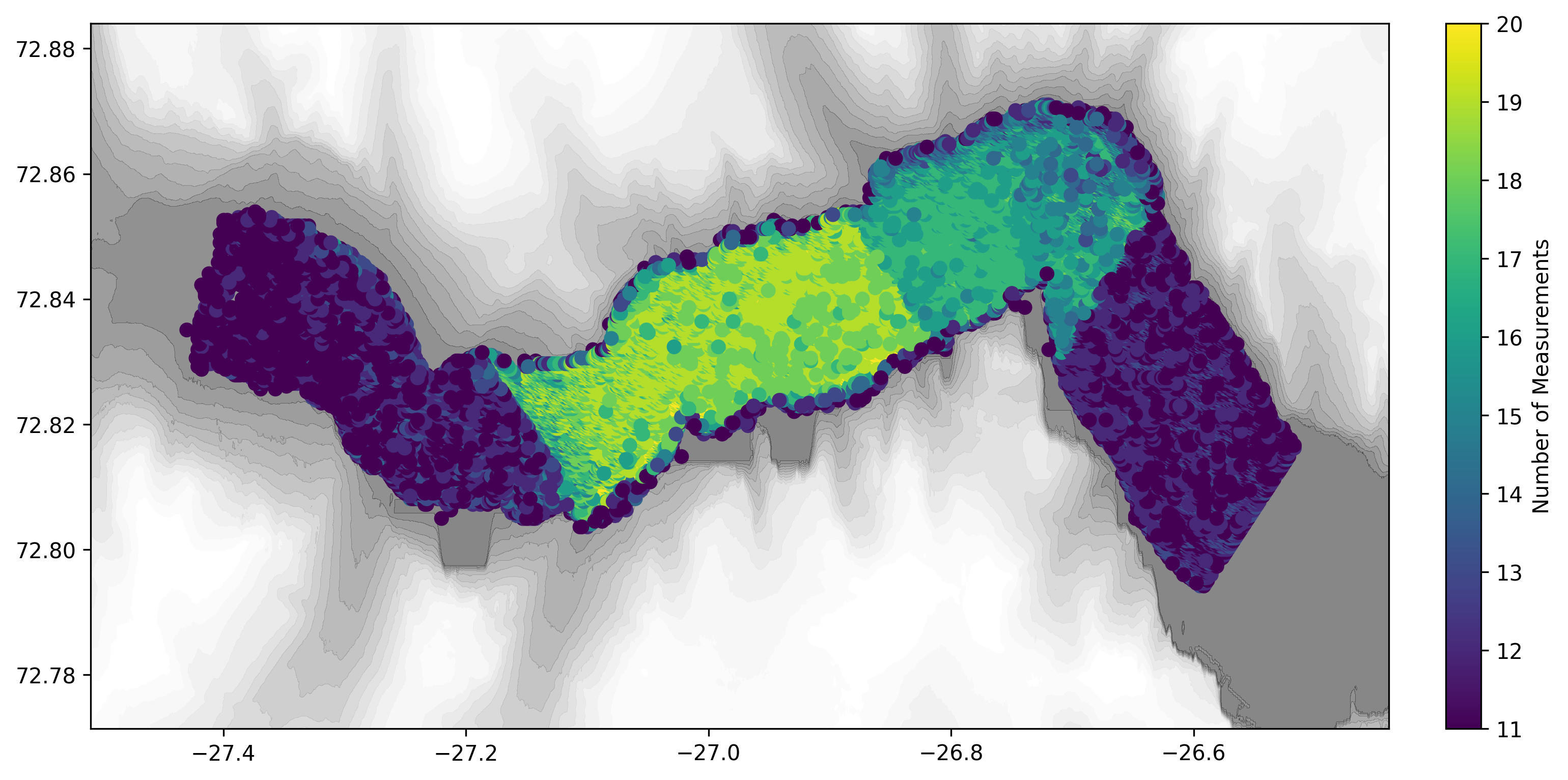} 
	\caption{\textbf{Number of SWOT measurements available for tidal estimation}.}
	\label{fig:num_measurements} 
\end{figure}

Due to the extreme sparsity of available SWOT data (less than 20 measurements over a full year), extreme care is needed when performing tidal harmonic analysis \cite{monahan2024tidal}. Harmonic analysis assumes tides can be described by the superposition of waves at discrete tidal frequencies. These frequencies exist at harmonics of the motions between the Earth, Moon, and Sun and are described as constituents. For a given constituent $k$, the corresponding tidal wave is given in quadrature by
\begin{equation}
    A_k \sin{\omega t} + B_k \cos{\omega t}.
\end{equation}
Comparisons of tidal constituents are done in terms of the amplitude $C_k = \sqrt{A_k^2 + B_k^2}$ and phase $\phi_k = \arctan{A_k / B_k}$. Modern tidal analysis is carried out in the time-domain using least-squares estimation, and can thus be applied to irregularly sampled time-series. To accomplish this we define the tidal estimation problem as a general linear model, such that the sea-level at any time is given by $y_i = w^\trp x_i + \epsilon_i$. Here, $y_i$ is the observed sea-level at time $i$, $x_i$ is the $i^{th}$ row of an $M \times N$ matrix of basis functions where $M$ is the number of measurements and $N = 2n +2$ with $n$ equal to the number of constituents, $w$ is a set of inferred weights, and $\epsilon_i$ is the non-tidal residual. From this, we define a design matrix $X$ given by
\begin{equation} \label{eq:1}
	X = [1, \sin{\omega_{0} t_i}, \cos{\omega_{0} t_i}, \dots, \sin{\omega_{k} t_i}, \cos{\omega_{k} t_i}]^\trp
\end{equation} 
where 1 corresponds to the bias, and the remaining values the quadrature amplitudes. The extreme sparsity of the Dickson Fjord reference series is such that only the dominant lunar tide, M2, can be estimated as shown in \cite{hart2024tides}. As such, the design matrix in \ref{eq:1} simplifies to   
\begin{equation} \label{eq:1}
	X = [1, \sin{\omega_{M2} t_i}, \cos{\omega_{M2} t_i}]^\trp.
\end{equation} 
where $\omega_{M2} =28.985$ \unit{\deg\per\hour}. Conventional harmonic analysis only considers measurements from a single spatial location. Due to the spatial coherence of the oceanic response to tidal forcing \cite{monahan2024response}, this procedure leaves considerable information out. SWOT data provides a complete picture of the instantaneous sea-surface height throughout the Dickson Fjord which can be exploited using an appropriate method. Here we adopt the spatially coherent harmonic analysis procedure in \cite{monahan2024tidal}. Readers are referred to \cite{monahan2024tidal} for a complete description of the procedure. However, the basic idea is to simultaneously estimate the quadrature amplitudes across a set of adjacent points by assuming that the amplitude at any point $P_{j,k}$ is given by the amplitude $w_{0,0}$ of the central point $P_{0,0}$ with a small offset denoted $w_{j,k}$. Our general linear model can be expanded to 
\begin{equation}
Y_{j,k} = X_{0,0}w_{0,0} + \rho X_{j,k}w_{j,k} 
\end{equation}
where $Y_{j,k}$ is the observation at point $P_{j,k}$, $X_{0,0}$ and $ X_{j,k}$ are the design matrices for points $P_{0,0}$ and $P_{j,k}$ respectively, and $\rho$ represents the probability that the observations $Y_{j,k}$ are correlated with $y_{0,0}$. By including the probability that the observations are correlated $\rho$, we impose an assumption that points with more similar time-series will have similar tides. This procedure is akin to a simple convolution of adjacent points with the central point. 
\\
\\
While the spatially coherent harmonic analysis can be used in tandem with any estimator, here we make use of the variational Bayesian (VB) estimator described above in S\ref{Bayes}. This choice is based on the following reasons. First, \cite{monahan2024tidal} find the VB approach to be less sensitive to both stationary (Gaussian) and non-stationary noise artifacts. Second, VB provides natural parameter shrinkage, which is helpful for reducing ``cross-talk'' between constituents left out of the analysis when only solving for M2. Lastly, the implicit uncertainty information is helpful in assessing the quality of the tidal estimates. The analysis shown in Figure \ref{fig:tides} only includes locations which have at least 15 observations. Figure \ref{fig:num_measurements} shows the distribution of SWOT measurements which can be used for tidal estimation through the fjord. A complete implementation of this approach, and code to replicate all tidal estimation is given in ``Fjord\_Tides.ipynb''. 

\subsubsection{Simple Analytical Seiche} \label{analytical}
In order to estimate the total seiche amplitude, and to relate the September and October events using the observed seismic observations, we consider a theoretical seiche acting on a simplified fjord geometry. We adopt the notation and fjord geometry employed in \cite{svennevig2024rockslide} for consistency. Here, we assume the seiche to act as an oscillating horizontal force directed N160$^\circ$E (perpendicular to the Dickson Fjord). Due to discrepancies between the defined geometries given in \cite{carrillo202416} and \cite{svennevig2024rockslide} we avoid prescribing values for the precise fjord dimensions except where necessary. The sloshing of the seiche produces a shift of the center of mass of the body of water $x$, and can be written as 
\begin{equation}
    x(t) = \Delta x \sin{\omega t}.
\end{equation}
Here $\Delta x$ is the amplitude of the horizontal oscillation and $\omega \approx 2\pi / 92$ \unit{\hertz} is the frequency of oscillation. As described in \cite{svennevig2024rockslide}, the amplitude of the Rayleigh waves produced is proportional to the magnitude of the horizontal force. Hence, we write $\Delta x$ in terms of the total force $F$. Taking the second derivative of the position of the center of mass we find
\begin{equation} \label{eq:force}
    F = \Delta x \omega^2 \sin{\omega t}.
\end{equation}
It can be seen that the maximum force occurs at the maximum displacement $\Delta x$ of the center of mass. Using SWOT data we can only observe the cross channel slope. As such, it is useful to relate the force back to the surface displacement $\Delta z$ such that 
\begin{equation}
    \Delta x = \frac{L}{3}\frac{\Delta z}{h + h_s}
\end{equation}
Equation \ref{eq:force} can be rewritten in terms of the vertical displacement $\Delta z$ as 
\begin{equation}
    F = \frac{L m \omega}{3} \frac{\Delta z}{h + h_s}
\end{equation}
We recognize that the surface displacement $\Delta z =  S L $ where $S$ is the cross-channel slope. Thus, the force $F$ can be written in terms of the cross-channel slope with 
\begin{equation}
    F = \frac{L^2 m \omega}{3} \frac{S}{h + h_s}.
\end{equation}
Thus, we have shown that the force is directly proportional to the cross-channel slope. This allows for the direct comparison between events. 

\subsubsection*{In-Situ Measurements}
In-situ measurements are provided by the CTD station located in the inner portion of the Dickson Fjord (as shown in Figure \ref{fig:overview} Panel A and can be accessed at \cite{Boone2023}). The station provides both standard meteorological and oceanic variables. Here we only make use of the wind speed and direction and water depth measurements. Data are sampled at 15-minute intervals which creates severe aliasing issues for observing the 92 \unit{\s} VLP signal. Due to the location of the device in the inner fjord, the seiche signal magnitude decreases beneath pre-event noise levels and is thus, unobservable in the data. 

\subsubsection{Dickson Fjord}
Here we present a brief overview of the Dickson Fjord. A more comprehensive description of the physiography and climate of the can be found in \cite{svennevig2024rockslide}. The Dickson Fjord sits at the terminus of the Hissinger Glacier in the northernmost area of the Kong Oscar Fjord system situated in East Greenland (See Figure \ref{fig:fjord_overview}). The Fjord itself sits deep in the Arctic Circle, and is thus characterized by sea-ice over much of the year. Sea ice dissipates in July and then forms again in October. The fjord fills a U-shaped valley basin, with multiple smaller glaciers situated on each side. The Fjord itself is 38 km long and between 2.5-3.2 km wide. The depth ranges from 150-200 \unit{\m} to 700 \unit{\m} from West to East with an approximate depth of 540 \unit{\m} in the center of the fjord across from the landslide location. Bathymetry estimates are taken from a 2018 survey by the National Danish Hydrographic Office at a resolution of 15 \unit{\m}. We note that no data exists between 150-300 \unit{m} meters of the coast due to the limitations of the survey vessel. This missing data creates large uncertainties in these regions which can significantly influence numerical simulations. 
\subsubsection{Tsunami Information}
\label{dickson_description}
\begin{figure} 
	\centering
	\includegraphics[width=1.0\textwidth]{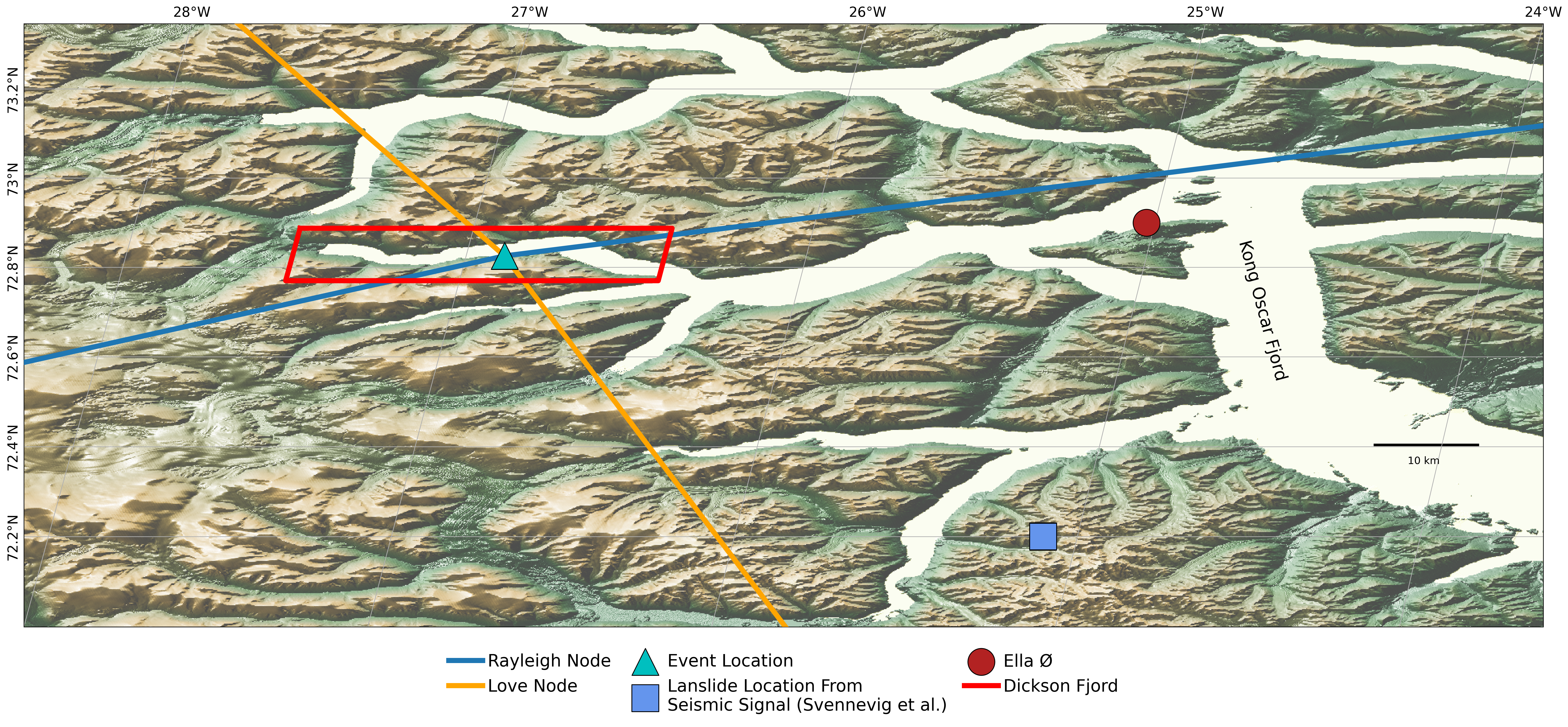} 
	\caption{\textbf{Overview map of the study region including relevant adjacent locations.} Background DEM is the Copernicus Global Digital Elevation Model \cite{ESA2024}.}
	\label{fig:fjord_overview} 
\end{figure}

Both tsunamis originated from landslides occurring in the same gully situated beneath an unnamed glacier \cite{svennevig2024rockslide}. These landslides were caused by debuttressing of the glacier following glacial thinning over the past decade. Direct observation of landslide scarring, and dirtying of the glacier using satellite imagery by both \cite{svennevig2024rockslide,carrillo202416} confirm this theory. Additionally, \cite{svennevig2024rockslide} evaluate the landslide dynamics of the September 16th landslide via seismic inversion. While the October 11th event did not produce a new landslide scar, a Sentinel-2 image showed significantly more erosion than was present after the September 16th event.
\\
\\
Empirical evidence of the two tsunamis is given by a combination of nearby in-situ sensors, and observed run-up height. Using satellite imagery, both \cite{svennevig2024rockslide, carrillo202416} observe an initially 200 \unit{\m} run-up height near the location of the slide, with an average of 60 \unit{\m} run-ups being observed through the remainder of the fjord for the September 16th event. Tsunami run-up for the October 11th even was only observed 200 \unit{\m} west of the gully approximately 75\% of the magnitude of the September event in this location (60 \unit{\m} vs 80 \unit{\m}). Almost 72 \unit{\km} at the Ella Ø research station, the initial run-up height was in excess of 4 \unit{\km} creating significant local damage. The location of Ella Ø relative to the Dickson Fjord is shown in Figure \ref{fig:fjord_overview}. To our knowledge, no information exists regarding the run-up at Ella Ø for the October 11th event due to arctic winter darkness. 

\bibliographystyle{unsrtnat}
\bibliography{references}  

\begin{thebibliography}{32}
\providecommand{\natexlab}[1]{#1}
\providecommand{\url}[1]{\texttt{#1}}
\expandafter\ifx\csname urlstyle\endcsname\relax
  \providecommand{\doi}[1]{doi: #1}\else
  \providecommand{\doi}{doi: \begingroup \urlstyle{rm}\Url}\fi

\bibitem[Diffenbaugh et~al.(2017)Diffenbaugh, Singh, Mankin, Horton, Swain, Touma, Charland, Liu, Haugen, Tsiang, et~al.]{diffenbaugh2017quantifying}
Noah~S Diffenbaugh, Deepti Singh, Justin~S Mankin, Daniel~E Horton, Daniel~L Swain, Danielle Touma, Allison Charland, Yunjie Liu, Matz Haugen, Michael Tsiang, et~al.
\newblock Quantifying the influence of global warming on unprecedented extreme climate events.
\newblock \emph{Proceedings of the National Academy of Sciences}, 114\penalty0 (19):\penalty0 4881--4886, 2017.

\bibitem[Overland(2022)]{overland2022arctic}
James~E Overland.
\newblock Arctic climate extremes.
\newblock \emph{Atmosphere}, 13\penalty0 (10):\penalty0 1670, 2022.

\bibitem[Landrum and Holland(2020)]{landrum2020extremes}
Laura Landrum and Marika~M Holland.
\newblock Extremes become routine in an emerging new arctic.
\newblock \emph{Nature Climate Change}, 10\penalty0 (12):\penalty0 1108--1115, 2020.

\bibitem[Carrillo-Ponce et~al.(2024)Carrillo-Ponce, Heimann, Petersen, Walter, Cesca, and Dahm]{carrillo202416}
Angela Carrillo-Ponce, Sebastian Heimann, Gesa~M Petersen, Thomas~R Walter, Simone Cesca, and Torsten Dahm.
\newblock The 16 september 2023 greenland megatsunami: Analysis and modeling of the source and a week-long, monochromatic seismic signal.
\newblock \emph{The Seismic Record}, 4\penalty0 (3):\penalty0 172--183, 2024.

\bibitem[Svennevig et~al.(2024)Svennevig, Hicks, Forbriger, Lecocq, Widmer-Schnidrig, Mangeney, Hibert, Korsgaard, Lucas, Satriano, et~al.]{svennevig2024rockslide}
Kristian Svennevig, Stephen~P Hicks, Thomas Forbriger, Thomas Lecocq, Rudolf Widmer-Schnidrig, Anne Mangeney, Cl{\'e}ment Hibert, Niels~J Korsgaard, Antoine Lucas, Claudio Satriano, et~al.
\newblock A rockslide-generated tsunami in a greenland fjord rang earth for 9 days.
\newblock \emph{Science}, 385\penalty0 (6714):\penalty0 1196--1205, 2024.

\bibitem[Rabinovich(2010)]{rabinovich2010seiches}
Alexander~B Rabinovich.
\newblock Seiches and harbor oscillations.
\newblock In \emph{Handbook of coastal and ocean engineering}, pages 193--236. World Scientific, 2010.

\bibitem[Amundson et~al.(2012)Amundson, Clinton, Fahnestock, Truffer, L{\"u}thi, and Motyka]{amundson2012observing}
Jason~M Amundson, John~F Clinton, Mark Fahnestock, Martin Truffer, Martin~P L{\"u}thi, and Roman~J Motyka.
\newblock Observing calving-generated ocean waves with coastal broadband seismometers, jakobshavn isbr{\ae}, greenland.
\newblock \emph{Annals of Glaciology}, 53\penalty0 (60):\penalty0 79--84, 2012.

\bibitem[of~Oceanography(1986)]{scripps1986global}
Scripps~Institution of~Oceanography.
\newblock Global seismograph network-iris/ida.
\newblock \emph{International Federation of Digital Seismograph Networks}, 1986.

\bibitem[(ASL)/USGS(1988)]{albuquerque1988global}
Albuquerque Seismological~Laboratory (ASL)/USGS.
\newblock Global seismograph network (gsn-iris/usgs).
\newblock \emph{International Federation of Digital Seismograph Networks}, 1988.

\bibitem[Mac{\'\i}as et~al.(2021)Mac{\'\i}as, Escalante, and Castro]{macias2021multilayer}
Jorge Mac{\'\i}as, Cipriano Escalante, and Manuel~J Castro.
\newblock Multilayer-hysea model validation for landslide-generated tsunamis--part 1: Rigid slides.
\newblock \emph{Natural Hazards and Earth System Sciences}, 21\penalty0 (2):\penalty0 775--789, 2021.

\bibitem[Chelton et~al.(2001)Chelton, Ries, Haines, Fu, and Callahan]{chelton2001satellite}
Dudley~B Chelton, John~C Ries, Bruce~J Haines, Lee-Lueng Fu, and Philip~S Callahan.
\newblock Satellite altimetry.
\newblock In \emph{International geophysics}, volume~69, pages 1--ii. Elsevier, 2001.

\bibitem[Abdalla et~al.(2021)Abdalla, Kolahchi, Ablain, Adusumilli, Bhowmick, Alou-Font, Amarouche, Andersen, Antich, Aouf, et~al.]{abdalla2021altimetry}
Saleh Abdalla, Abdolnabi~Abdeh Kolahchi, Micha{\"e}l Ablain, Susheel Adusumilli, Suchandra~Aich Bhowmick, Eva Alou-Font, Laiba Amarouche, Ole~Baltazar Andersen, Helena Antich, Lotfi Aouf, et~al.
\newblock Altimetry for the future: Building on 25 years of progress.
\newblock \emph{Advances in Space Research}, 68\penalty0 (2):\penalty0 319--363, 2021.

\bibitem[Morrow et~al.(2019)Morrow, Fu, Ardhuin, Benkiran, Chapron, Cosme, d’Ovidio, Farrar, Gille, Lapeyre, et~al.]{morrow2019global}
Rosemary Morrow, Lee-Lueng Fu, Fabrice Ardhuin, Mounir Benkiran, Bertrand Chapron, Emmanuel Cosme, Francesco d’Ovidio, J~Thomas Farrar, Sarah~T Gille, Guillaume Lapeyre, et~al.
\newblock Global observations of fine-scale ocean surface topography with the surface water and ocean topography (swot) mission.
\newblock \emph{Frontiers in Marine Science}, 6:\penalty0 232, 2019.

\bibitem[Fj{\o}rtoft et~al.(2013)Fj{\o}rtoft, Gaudin, Pourthi{\'e}, Lalaurie, Mallet, Nouvel, Martinot-Lagarde, Oriot, Borderies, Ruiz, et~al.]{fjortoft2013karin}
Roger Fj{\o}rtoft, Jean-Marc Gaudin, Nadine Pourthi{\'e}, Jean-Claude Lalaurie, Alain Mallet, Jean-Fran{\c{c}}ois Nouvel, Joseph Martinot-Lagarde, Helene Oriot, Pierre Borderies, Christian Ruiz, et~al.
\newblock Karin on swot: Characteristics of near-nadir ka-band interferometric sar imagery.
\newblock \emph{IEEE Transactions on Geoscience and Remote Sensing}, 52\penalty0 (4):\penalty0 2172--2185, 2013.

\bibitem[Pasyanos et~al.(2014)Pasyanos, Masters, Laske, and Ma]{pasyanos2014litho1}
Michael~E Pasyanos, T~Guy Masters, Gabi Laske, and Zhitu Ma.
\newblock Litho1. 0: An updated crust and lithospheric model of the earth.
\newblock \emph{Journal of Geophysical Research: Solid Earth}, 119\penalty0 (3):\penalty0 2153--2173, 2014.

\bibitem[Gelman et~al.(2019)Gelman, Goodrich, Gabry, and Vehtari]{gelman2019r}
Andrew Gelman, Ben Goodrich, Jonah Gabry, and Aki Vehtari.
\newblock R-squared for bayesian regression models.
\newblock \emph{The American Statistician}, 2019.

\bibitem[Caceres et~al.(2003)Caceres, Valle-Levinson, and Atkinson]{caceres2003observations}
Mario Caceres, Arnoldo Valle-Levinson, and Larry Atkinson.
\newblock Observations of cross-channel structure of flow in an energetic tidal channel.
\newblock \emph{Journal of Geophysical Research: Oceans}, 108\penalty0 (C4), 2003.

\bibitem[Cottier et~al.(2010)Cottier, Nilsen, Skogseth, Tverberg, Skar{\dh}hamar, and Svendsen]{cottier2010arctic}
FR~Cottier, Frank Nilsen, R~Skogseth, Vigdis Tverberg, J~Skar{\dh}hamar, and H~Svendsen.
\newblock Arctic fjords: a review of the oceanographic environment and dominant physical processes.
\newblock \emph{Geological Society, London, Special Publications}, 344\penalty0 (1):\penalty0 35--50, 2010.

\bibitem[Hart-Davis et~al.(2024)Hart-Davis, Andersen, Ray, Zaron, Schwatke, Arildsen, and Dettmering]{hart2024tides}
MG~Hart-Davis, OB~Andersen, RD~Ray, ED~Zaron, C~Schwatke, RL~Arildsen, and D~Dettmering.
\newblock Tides in complex coastal regions: early case studies from wide-swath swot measurements.
\newblock \emph{Authorea Preprints}, 2024.

\bibitem[Stammer et~al.(2014)Stammer, Ray, Andersen, Arbic, Bosch, Carrere, Cheng, Chinn, Dushaw, Egbert, et~al.]{stammer2014accuracy}
Detlef Stammer, Richard~D Ray, Ole~Baltazar Andersen, Brian~K Arbic, Wolfgang Bosch, Laurent Carrere, Yongcun Cheng, Douglas~S Chinn, Brian~D Dushaw, Gary~D Egbert, et~al.
\newblock Accuracy assessment of global barotropic ocean tide models.
\newblock \emph{Reviews of Geophysics}, 52\penalty0 (3):\penalty0 243--282, 2014.

\bibitem[Monahan et~al.(2024{\natexlab{a}})Monahan, Tang, Roberts, and Adcock]{monahan2024tidal}
Thomas Monahan, Tianning Tang, Stephen Roberts, and Thomas~AA Adcock.
\newblock Tidal and mean sea surface corrections from and for swot using a spatially coherent variational bayesian harmonic analysis.
\newblock \emph{Authorea Preprints}, 2024{\natexlab{a}}.

\bibitem[Price et~al.(1987)Price, Weller, and Schudlich]{price1987wind}
James~F Price, Robert~A Weller, and Rebecca~R Schudlich.
\newblock Wind-driven ocean currents and ekman transport.
\newblock \emph{Science}, 238\penalty0 (4833):\penalty0 1534--1538, 1987.

\bibitem[Rantanen et~al.(2022)Rantanen, Karpechko, Lipponen, Nordling, Hyv{\"a}rinen, Ruosteenoja, Vihma, and Laaksonen]{rantanen2022arctic}
Mika Rantanen, Alexey~Yu Karpechko, Antti Lipponen, Kalle Nordling, Otto Hyv{\"a}rinen, Kimmo Ruosteenoja, Timo Vihma, and Ari Laaksonen.
\newblock The arctic has warmed nearly four times faster than the globe since 1979.
\newblock \emph{Communications earth \& environment}, 3\penalty0 (1):\penalty0 168, 2022.

\bibitem[Krischer et~al.(2015)Krischer, Megies, Barsch, Beyreuther, Lecocq, Caudron, and Wassermann]{krischer2015obspy}
Lion Krischer, Tobias Megies, Robert Barsch, Moritz Beyreuther, Thomas Lecocq, Corentin Caudron, and Joachim Wassermann.
\newblock Obspy: A bridge for seismology into the scientific python ecosystem.
\newblock \emph{Computational Science \& Discovery}, 8\penalty0 (1):\penalty0 014003, 2015.

\bibitem[Xu and Mikesell(2022)]{xu2022estimation}
Zongbo Xu and T~Dylan Mikesell.
\newblock Estimation of resolution and covariance of ambient seismic source distributions: Full waveform inversion and matched field processing.
\newblock \emph{Journal of Geophysical Research: Solid Earth}, 127\penalty0 (6):\penalty0 e2022JB024374, 2022.

\bibitem[Fox and Roberts(2012)]{fox2012tutorial}
Charles~W Fox and Stephen~J Roberts.
\newblock A tutorial on variational bayesian inference.
\newblock \emph{Artificial intelligence review}, 38:\penalty0 85--95, 2012.

\bibitem[Roberts et~al.(2013)Roberts, McQuillan, Reece, and Aigrain]{roberts2013astrophysically}
S~Roberts, A~McQuillan, S~Reece, and S~Aigrain.
\newblock Astrophysically robust systematics removal using variational inference: application to the first month of kepler data.
\newblock \emph{Monthly Notices of the Royal Astronomical Society}, 435\penalty0 (4):\penalty0 3639--3653, 2013.

\bibitem[Ruanaidh and Fitzgerald(2012)]{ruanaidh2012numerical}
Joseph JK~O Ruanaidh and William~J Fitzgerald.
\newblock \emph{Numerical Bayesian methods applied to signal processing}.
\newblock Springer Science \& Business Media, 2012.

\bibitem[Penny and Roberts(2002)]{penny2002bayesian}
WD~Penny and SJ~Roberts.
\newblock Bayesian multivariate autoregressive models with structured priors.
\newblock \emph{IEE Proceedings-Vision, Image and Signal Processing}, 149\penalty0 (1):\penalty0 33--41, 2002.

\bibitem[Monahan et~al.(2024{\natexlab{b}})Monahan, Tang, Roberts, and Adcock]{monahan2024response}
Thomas Monahan, Tianning Tang, Stephen Roberts, and Thomas Adcock.
\newblock Response framework: Tidal analysis and prediction through physics-informed ml.
\newblock 2024{\natexlab{b}}.

\bibitem[Boone et~al.(2023)Boone, Rysgaard, Frandsen, Develter, Institute, and University]{Boone2023}
W.~Boone, S.~Rysgaard, E.R. Frandsen, R.~Develter, Flanders~Marine Institute, and Aarhus University.
\newblock Greenland integrated observatory - ctd \& atmospheric station dickson fjord - 2023.
\newblock \url{https://doi.org/10.14284/637}, 2023.

\bibitem[{European Space Agency}(2024)]{ESA2024}
{European Space Agency}.
\newblock Copernicus global digital elevation model, 2024.
\newblock URL \url{https://doi.org/10.5069/G9028PQB}.
\newblock Distributed by OpenTopography. Accessed: 2024-10-10.

\end{thebibliography}






\section*{Additional Figures}\label{secA1}

\begin{table} 
	\centering
	\caption{\textbf{Bounding box for SWOT querying}}
	\label{tab:bounding_box} 
	\begin{tabular}{lcr} 
		\\
		\hline
		  Corner & Lat & Lon\\
		\hline
		Bottom Left & 72.77 & -27.51 \\
		Top Left & 72.89 & -27.51 \\
		Bottom Right & 72.77 & -26.42 \\
            Top Right & 72.89 & -26.42 \\
		\hline
	\end{tabular}
\end{table}
\end{document}